# Emotion Recognition Using Speaker Cues

Ismail Shahin
Department of Electrical and Computer Engineering
University of Sharjah
Sharjah, United Arab Emirates
ismail@sharjah.ac.ae

*Abstract*— This research aims at identifying the unknown emotion using speaker cues. In this study, we identify the unknown emotion using a two-stage framework. The first stage focuses on identifying the speaker who uttered the unknown emotion, while the next stage focuses on identifying the unknown emotion uttered by the recognized speaker in the prior stage. This proposed framework has been evaluated on an Arabic Emirati-accented speech database uttered by fifteen speakers per gender. "Mel-Frequency Cepstral Coefficients (MFCCs) have been used as the extracted features and Hidden Markov Model (HMM)" has been utilized as the classifier in this work. Our findings demonstrate that emotion recognition accuracy based on the two-stage framework is greater than that based on the one-stage approach and the "state-of-the-art classifiers and models such as Gaussian Mixture Model (GMM), Support Vector Machine (SVM), and Vector Quantization (VQ)". The average emotion recognition accuracy based on the two-stage approach is 67.5%, while the accuracy reaches to 61.4%, 63.3%, 64.5%, and 61.5%, based on the one-stage approach, GMM, SVM, and VQ, respectively. The achieved results based on the two-stage framework are very close to those attained in subjective assessment by human listeners.

## I. INTRODUCTION

Emotion recognition is made up of two elements: emotion identification and emotion verification. In emotion identification, a speech sample from the unknown emotion is analyzed and competed with models of known emotions. The unknown emotion is recognized as the emotion whose model best fits the input speech signal. In emotion verification, the aim is to decide whether an emotion corresponds to a certain known emotion or to some other unknown emotions [1].

Emotion recognition can be utilized in many various applications in "telecommunications, human robotic interfaces, smart call centers, and intelligent spoken tutoring systems. In telecommunications, emotion recognition emerges in assessing the emotion of a caller for telephone response facilities. Recognizing emotions can also be exploited in human robotic interfaces where robots can be trained to interact with humans and recognize human emotions. Further applications can be observed in smart call centers where probable problems happening from poor communications can be sensed by emotion recognition. Emotion recognition can be used in intelligent verbal tutoring to sense and alter to the emotions of students when students ran into a tedious situation throughout tutoring gatherings" [2], [3], [4].

Emotion recognition typically performs in one of two input forms, "text-dependent form or text-independent form. In the text-dependent form, emotions supply utterances of the alike text for both training and testing assessments. In the text-independent form, emotions are not limited to give specific texts in evaluation trials".

In this work, a two-stage architecture has been proposed, applied, and tested to identify the unknown emotion. The first stage identifies the unknown speaker followed by the second stage that identifies the unknown emotion which is uttered by the recognized speaker in the first stage.

This paper is organized as follows: Section II covers the literature review of emotion recognition. Section III explains the used dataset and extraction of features. Section IV details the proposed framework of emotion recognition and the experiments. Section V discusses the experiments and the achieved results. Finally, section VI gives conclusion of this work.

## II. LITERATURE REVIEW

In the last three decades, emotion recognition becomes a hot topic in the research community. Emotion recognition accuracy is still imperfect. There are many studies that focus on enhancing emotion recognition accuracy using English and Arabic Corpora [1, 5-14].

Emotion recognition has been investigated in many studies [1, 5-11]. Yogesh *et.al* [5] introduced "a novel particle swarm optimization aided biogeography-based method for feature selection. They did their experiments utilizing Berlin Emotional Speech corpus (BES), Surrey Audio-Visual Expressed Emotion corpus (SAVEE), and Speech Under Simulated and Actual Stress (SUSAS) corpus". Shahin shed the light in one of his prior work [6] on studying and improving "text-independent and speaker-independent talking condition identification in stressful and emotional environments" (completely two separate environments) based on three independent and distinct classifiers: "Hidden Markov Models (HMMs), Second-Order Circular Hidden Markov Models (CHMM2s), and Supra-segmental Hidden Markov Models (SPHMMs)". His study demonstrated that "SPHMMs lead HMMs and CHMM2s for emotion recognition in the two environments" [6]. Shahin and Ba-Hutair [7] investigated in one of their investigations on how to enhance "text-independent and speaker-independent talking condition recognition in each of stressful and emotional environments based on Second-Order Circular Suprasegmental Hidden Markov Models (CSPHMM2s) as classifiers". In addition, one of the main targets of their research was to distinguish "stressful talking environment from emotional talking environment based on CSPHMM2s. They arrived at a judgement that talking recognition in stressful and emotional environments based on CSPHMM2s is a leader to that based on HMMs, CHMM2s, and SPHMMs". In one of his previous work [8], Shahin utilized emotions to recognize the unknown speakers. He proposed a novel approach to recognize speakers from their emotions based on HMMs. In another study by Shahin [9], he proposed, implemented, and assessed "speaker-dependent and text-dependent speaking style authentication (verification) systems that accept or reject the identity claim of a speaking style based on SPHMMs". His findings, based on SPHMMs, demonstrated that the "average speaking style authentication performance is: 99%, 37%, 85%, 60%, 61%, 59%, 41%, 61%, and 57% belonging, respectively, to the speaking styles: neutral, shouted, slow, loud, soft, fast, angry, happy, and fearful". To improve "speaker verification accuracy in emotional talking environments, Shahin [10] introduced a two-stage approach that employs the emotion of speaker cues (text-independent and emotion-dependent speaker verification problem) based on both HMMs and SPHMMs as classifiers". His architecture is made up of "two sequential stages that combine and integrate emotion recognizer followed by a speaker recognizer into one recognizer". His approach has been tested on two different and independent emotional speech corpora: his captured corpus and "Emotional Prosody Speech and Transcripts (EPST)" corpus. The results of his study showed that his framework gave better results with a significant improvement over prior work and other approaches such as "emotion-independent speaker verification approach and emotion-dependent speaker verification approach based entirely on HMMs". In one more research by Shahin and Bou Nassif [11], they targeted to improve emotion recognition accuracy based on "Third-Order Hidden Markov Models (HMM3s) as a classifier". Their work has been evaluated on EPST dataset. "The extracted features of EPST database are Mel-Frequency Cepstral Coefficients (MFCCs)". Their findings yielded an average emotion recognition accuracy of 71.8%. Shahin [1] spotlighted on recognizing the unknown emotion based on the "Third-Order Circular Suprasegmental Hidden Markov Model (CSPHMM3) as a classifier". His research has been assessed on EPST dataset. The extracted features of EPST database are the MFCCs. His results gave average emotion recognition accuracy of 77.8% based on the CSPHMM3. The results of his study showed that "CSPHMM3 is superior to HMM3, Gaussian Mixture Model (GMM), Support Vector Machine (SVM), and Vector Quantization (VQ) by 6.0%, 4.9%, 3.5%, and 5.4%, respectively, for emotion recognition" [1].

There are not many studies that focus on emotion recognition using Arabic speech databases [12-14]. Klaylat et. al [12] introduced a two-phase model to enhance emotion recognition system. Their system recognizes three different emotions: happy, angry, and surprised utilizing an Arabic speech database. They implemented thirty five classification models and a Sequential Minimal Optimization (SMO) classifier in their work. Their results yielded 95.52% as an emotion recognition accuracy [12]. El Gohary et. al [13] concerned with emotion detection in Arabic text. Their study is based on a moderate sized Arabic emotion lexicon used to annotate Arabic children stories for six distinct emotions: anger, fear, joy, surprise, sadness, and disgust. They attained 65% as an emotion detection accuracy in Arabic text. Shahin et. al [14] focused on "recognizing text-independent and speaker-independent emotions using Arabic Emirati-accented speech dataset based on a proposed hybrid classifier called cascaded Gaussian Mixture Model and Deep Neural Network, GMM-DNN, (GMM followed by DNN). Six diverse emotions have been utilized in their study. These emotions are: neutrality, happiness, sadness, disgust, anger, and fear. They achieved 83.97% as an average emotion recognition accuracy using the novel GMM-DNN classifier" [14].

In this research, we propose, implement, and assess a text-independent two-stage emotion recognizer based on Hidden Markov Model (HMM) as a classifier in each stage. The first stage is nothing but a speaker recognizer followed by an emotion recognizer. Our work has been tested on our captured Arabic Emirati-accented speech database that is made up of fifteen Emirati speakers per gender talking in six different emotions using MFCCs as the extracted features.

III. SPEECH DATASET AND EXTRACTION OF FEATURES

A. Speech Dataset

In this work, our proposed approach has been evaluated on an "Arabic Emirati-accented speech dataset captured from fifteen local Emirati speakers per gender. Speakers utter eight Emirati sentences that are regularly used in the UAE society. Each sentence is uttered by every speaker nine times under each of neutral, happy, sad, disgust, angry, and fear emotions. Table 1 exhibits the database used in this research where the right column shows the utterances in Emirati dialect while the left column exhibits the English translation of these sentences. This database was picked up in two separated and diverse sessions: training session and testing session. The dataset was recorded in an uncontaminated environment in the College of Communication, University of Sharjah, United Arab Emirates by a set of skilled engineering students. The database was recorded by a speech acquisition board using a 16-bit linear coding A/D converter and sampled at a sampling rate of 44.6 kHz".

B. Extraction of Features

Static Mel-Frequency Cepstral Coefficients (static MFCCs) and delta MFCCs define the phonetic content of speech signals in our corpus. Such coefficients have been mainly used in many studies of emotion recognition and speaker recognition [7], [10], [14], [15], [16], [17], [18]. The observation vector in HMM has been expressed by the MFCC feature analysis. A 32-dimension MFCC (16 static MFCCs and 16 delta MFCCs) feature analysis forms the observation vectors in a 6-state HMM model with a continuous mixture observation density.

IV. PROPOSED TWO-STAGE EMOTION RECOGNTION APPROACH AND THE EXPERIMENTS

Our proposed two-stage emotion recognizer is given in Fig. 1. This recognizer is comprised of two cascaded recognizers:

First Stage: Speaker Recognizer

Based on our proposed approach, the first stage is to identify the unknown speaker who produced the utterance in the unknown emotion (speaker identification problem). In this stage, there are $n$ probabilities that are computed based on HMM and the highest probability is chosen as the recognized speaker as shown in the following formula,

$$S^* = arg\ max\ 1_{n \geq s \geq 1}\ \{P(O|\lambda^s)\} \quad (1)$$

where, "$S^*$ is the index of the recognized speaker, $O$ is the observation sequence of the unknown speaker who uttered the utterance in an unknown emotion, and $P(O|\lambda^s)$ is the probability of the observation sequence $O$ of the unknown speaker given the $s$th HMM speaker model".

In this stage, the "$s$th HMM speaker model has been derived in the training session for every speaker talking in neutral condition using all the first four sentences of the database with a repetition of nine utterances/sentence. The overall number of utterances used to build every HMM speaker model in this session is 36 (4 sentences × 9 utterances/sentence)".

Second Stage: Emotion Recognizer

The second stage of our proposed framework is to identify the unknown emotion given that the unknown speaker was recognized in the first stage (speaker-specific emotion identification problem). In this stage, $m$ probabilities per speaker are calculated based on HMM and the maximum probability is chosen as the identified emotion for the recognized speaker as given in the following formula,

$$E^* = arg\ max\ 1_{m \geq e \geq 1}\ \{P(e|\lambda^e, S^*)\} \quad (2)$$

where, "$E^*$ is the index of the identified emotion, $P(e|\lambda^e, S^*)$ is the probability of the observation sequence $e$ that corresponds to the unknown emotion given the $e$th HMM emotion model and the identified speaker".

The "$e$th HMM emotion model for every speaker has been obtained using nine utterances per sentence (the first four sentences of the corpus). The total number of utterances utilized to get every speaker-dependent HMM emotion model is 36 (4 sentences × 9 utterances/sentence)".

In the "test or identification session (completely independent from the training session), each one of the thirty speakers used nine utterances per sentence (last four sentences of the dataset) under each emotion including the neutral state. The whole number of utterances utilized in this session is 6,480 (30 speakers × 4 sentences × 9 utterances/sentence × 6 emotions)".

## V. RESULTS AND DISCUSSION

In this work, a two-stage approach has been proposed, applied, and tested to enhance emotion recognition performance using our captured Arabic Emirati-accented dataset. MFCC has been used as the extracted features and HMM has been used as the adopted classifier. We evaluated our proposed architecture using six different emotions spoken by fifteen speakers per gender. The six emotions are: "neutral, happy, sad, disgust, angry, and fear".

Table II shows emotion recognition accuracy using the one-stage framework (identifying the unknown emotion directly without identifying the unknown speaker). This table yields average emotion recognition accuracy of 61.4%. It is evident from this table that the highest emotion identification accuracy happens when speakers speak neutrally, while the least accuracy takes place when speakers speak angrily.

Based on the two-stage approach, the average speaker identification accuracy in the first stage is 72.8%. The emotion identification accuracy in the second stage of our proposed approach is given in Table III. This table gives average emotion identification accuracy of 67.5%. This table apparently demonstrates that the maximum accuracy occurs when speakers talk neutrally, whereas the minimum accuracy happens when speakers communicate angrily.

To conform whether "emotion recognition accuracy differences (emotion recognition accuracy based on the two-stage approach and that based on the one-stage approach) are actual or just come from statistical differences, a statistical significance test has been performed. The statistical significance test has been done based on the Student's $t$ Distribution test" as given by,

$$t_{model\ x, model\ y} = \frac{\bar{x}_{model\ x} - \bar{x}_{model\ y}}{SD_{pooled}} \quad (3)$$

where "$\bar{x}_{model\ x}$ is the mean of the first sample (model $x$) of size $n$, $\bar{x}_{model\ y}$ is the mean of the second sample (model $y$) of equal size, and $SD_{pooled}$ is the pooled standard deviation of the two samples (models $x$ and $y$)" given as,

$$SD_{pooled} = \sqrt{\frac{SD_{model\ x}^2 + SD_{model\ y}^2}{2}} \quad (4)$$

where "$SD_{model\ x}$ is an estimate of the standard deviation of the average of the first sample (model $x$) of size $n$ and $SD_{model\ y}$ is an estimate of the standard deviation of the average of the second sample (model $y$) of same size".

The "computed $t$ value between the two-stage architecture and the one-stage framework is calculated based on Table II and Table III. The calculated value is $t_{two-stage,\ one-stage}$ = 1.798. This calculated value is greater than the tabulated critical value $t_{0.05}$ = 1.645 at 0.05 significant level". Hence, it is evident that the two-stage approach yields higher emotion recognition accuracy than the one-stage approach.

Emotion recognition accuracy based on the two-stage framework has been compared with that based on the "state-of-the-art classifiers and models such as Gaussian Mixture Model (GMM) [19], Support Vector Machine (SVM) [20], and Vector Quantization (VQ)" [21]. The average emotion recognition accuracy based on "GMM, SVM, and VQ" is 63.3%, 64.5%, and 61.5%, respectively. It is apparent from this experiment that the two-stage approach gives greater emotion recognition accuracy than each one of these three classifiers and models.

An "informal subjective assessment for emotion recognition using our collected corpus has been carried out using 10 human non-professional listeners. In this assessment, a total of 540 utterances (30 speakers × 6 emotions × 3 repetitions) have been involved. These listeners are asked two questions. The first one is to recognize the speaker and the second question is to recognize the unknown emotion given that the speaker was already recognized. Based on this assessment, the average speaker recognition accuracy is 70.4% and the average emotion recognition accuracy is 65.4%. This average emotion recognition accuracy is very similar to the achieved average based on the two-stage architecture (67.5%)".

## VI. Concluding Remarks

In this work, a two-stage emotion recognizer that is composed of a speaker recognizer followed by an emotion recognizer has been introduced, applied, and assessed on an "Arabic Emirati-accented speech dataset". Some concluding remarks can be drawn in this study. First, the two-stage emotion recognizer gives better accuracy than the one-stage emotion recognizer. Therefore, speaker cues help enhancing emotion recognition accuracy. Second, the two-stage emotion recognizer leads the "state-of-the-art classifiers and models such as GMM, SVM, and VQ". Third, the highest emotion recognition accuracy occurs when speakers speak neutrally. Finally, the least emotion recognition accuracy takes place when speakers talk angrily.

This work has some limitations. Firstly, our dataset is limited to six emotions only. Secondly, the attained emotion recognition accuracy based on the two-stage framework is imperfect. This is because the accuracy of the two-stage approach is the resultant of two accuracies:

a) Speaker recognition accuracy which is nonideal.
b) Emotion recognition accuracy which is imperfect.

Our coming plan is to use "Deep Neural Network (DNN)" to obtain better accuracy [22]. Furthermore, our strategy is to analyze Emirati-accented emotion recognition in biased talking environments [23], [24].

## Acknowledgment

The author of this work wishes to express his thanks to the "University of Sharjah for funding this research through the competitive research project entitled Emirati-Accented Speaker and Emotion Recognition Based on Deep Neural Network, No. 19020403139".

## References

[1] I. Shahin, "Emotion recognition based on third-order circular suprasegmental hidden Markov model," The 2019 IEEE Jordan International Joint Conference on Electrical Engineering and Information Technology (JEEIT), April 2019, Amman, Jordan, pp. 599-604.

[2] V.A. Petrushin, "Emotion recognition in speech signal: experimental study, development, and application," Proceedings of International Conference on Spoken Language Processing, ICSLP 2000.

[3] R. Cowie, E. Douglas-Cowie, N. Tsapatsoulis, G. Votsis S. Collias, W. Fellenz, and J. Taylor, "Emotion recognition in human-computer interaction," IEEE Signal Processing Magazine, 18 (1), 2001, pp. 32-80.

[4] N. Fragopanagos and J. G. Taylor, "Emotion recognition in human-computer interaction," Neural Networks, special issue (18), 2005, pp. 389-405.

[5] C. K. Yogesh, M. Hariharan, R. Ngadiran, A. H. Adom, S. Yaacob, C. Berkai, and K. Polat, "A new hybrid PSO assisted biogeography-based optimization for emotion and stress recognition from speech signal," Expert Systems with Applications, Vol. 69, 2017, pp.149-158.

[6] I. Shahin, "Studying and enhancing talking condition recognition in stressful and emotional talking environments based on HMMs, CHMM2s and SPHMMs," Journal on Multimodal User Interfaces, Vol. 6, issue 1, June 2012, pp. 59-71, DOI: 10.1007/s12193-011-0082-4.

[7] I. Shahin and Mohammed Nasser Ba-Hutair, "Talking condition recognition in stressful and emotional talking environments based on CSPHMM2s," International Journal of Speech Technology, Vol. 18, issue 1, March 2015, pp. 77-90, DOI: 10.1007/s10772-014-9251-7.

[8] I. Shahin, "Using emotions to identify speakers," The 5th International Workshop on Signal Processing and its Applications (WoSPA 2008), Sharjah, United Arab Emirates, March 2008.

[9] I. Shahin, "Speaking style authentication using suprasegmental hidden Markov models," University of Sharjah Journal of Pure and Applied Sciences, Vol. 5, No. 2, June 2008, pp. 41-65.

[10] I. Shahin, "Employing emotion cues to verify speakers in emotional talking environments," Journal of Intelligent Systems, Special Issue on Intelligent Healthcare Systems, DOI: 10.1515/jisys-2014-0118, Vol. 25, issue 1, January 2016, pp. 3-17.

[11] I. Shahin and Ali Bou Nassif, "Utilizing third-order hidden Markov models for emotional talking condition recognition," 14th IEEE International Conference on Signal Processing (ICSP2018) 12-16 August 2018, Beijing, China, pp. 250-254.

[12] S. Klaylat, Z. Osman, L. Hamandi, and R. Zantout, "Enhancement of an Arabic speech emotion recognition system," International Journal of Applied Engineering Research, Vol. 13, issue 5, 2018, pp. 2380–2389.

[13] A. F. El-Gohary, T. I. Sultan, M. A. Hana, and M. M. El Dosoky, "A Computational approach for analyzing and detecting emotions in Arabic text," International Journal of Engineering Research and Applications (IJERA), 2013, Vol. 3, pp. 100-107.

[14] I. Shahin, Ali Bou Nassif, and Shibani Hamsa, "Emotion Recognition using Hybrid Gaussian Mixture Model and Deep Neural Network," IEEE Access, Vol. 7, March 2019, pp. 26777 - 26787, DOI 10.1109/ACCESS.2019.2901352.

[15] I. Shahin, "Employing second-order circular suprasegmental hidden Markov models to enhance speaker identification performance in shouted talking environments," EURASIP Journal on Audio, Speech, and Music Processing, Vol. 2010, Article ID 862138, 10 pages, June 2010. doi:10.1155/2010/862138.

[16] I. Shahin, "Speaker identification in a shouted talking environment based on novel third-order circular suprasegmental hidden Markov models," Circuits, Systems and Signal Processing, DOI: 10.1007/s00034-015-0220-4, Vol. 35, issue 10, October 2016, pp. 3770-3792.

[17] I. Shahin, "Employing both gender and emotion cues to enhance speaker identification performance in emotional talking environments," International Journal of Speech Technology, Vol. 16, issue 3, September 2013, pp. 341-351, DOI: 10.1007/s10772-013-9188-2.

[18] D. Morrison, R. Wang, and L. C. De Silva, "Ensemble methods for spoken emotion recognition in call-centres," Speech Communication, Vol. 49, issue 2, February 2007, pp. 98-112.

[19] D. A. Reynolds, "Speaker identification and verification using Gaussian mixture speaker models," Speech Communication, Vol. 17, 1995, pp. 91-108.

[20] V. Wan and W. M. Campbell, "Support vector machines for speaker verification and identification," Neural Networks for Signal Processing X, Proceedings of the 2000 IEEE Signal Processing Workshop, Vol. 2, 2000, pp. 775 – 784.

[21] T. Kinnunen and H. Li, "An overview of text-independent speaker recognition: From features to supervectors," Speech Communication, Vol. 52, No. 1, January 2010, pp. 12 – 40.

[22] A. B. Nassif, I. Shahin, I. Attili, M. Azzeh, and K. Shaalan, "Speech recognition using deep neural networks: a Systematic Review," IEEE Access, Vol. 7, February 2019, pp. 19143 - 19165, DOI: 10.1109/ACCESS.2019.2896880.

[23] I. Shahin, "Speaker identification investigation and analysis in unbiased and biased emotional talking environments," International Journal of Speech Technology, Vol. 15, issue 3, September 2012, pp. 325-334, DOI: 10.1007/s10772-012-9156-2.

[24] I. Shahin, "'Analysis and investigation of emotion identification in biased emotional talking environments," IET Signal Processing, Vol. 5, No. 5, August 2011, pp. 461 – 470, DOI: 10.1049/iet-spr.2010.0059.

TABLE I.  EMIRATI DATASET AND ITS ENGLISH VERSION.

| No. | English Version | Emirati Accent |
|---|---|---|
| 1. | I'm leaving now, may God keep you safe. | فداعة الرحمن بترخص عنكم الحينه. |
| 2. | The one whose hand is in the water is not the same as he/she whose hand is in fire. | اللي ايده في الماي مب نفس اللي ايده في الضو. |
| 3. | Where do you want to go today? | وين تبون تسيرون اليوم؟ |
| 4. | The weather is nice, let's sit outdoors. | قوموا نيلس في الحوي . الجوغاوي برع. |
| 5. | What's in the pot, the spoon gets out. | اللي في الجدر يطلعه الملاس. |
| 6. | Welcome millions, and they are not enough. | مرحبا ملايين ولا يسدن. |
| 7. | Get ready, I will pick you up tomorrow. | زهب عمرك بخطف علياك باجر. |
| 8. | He/she who doesn't know the value of the falcon, will grill it. | اللي ما يعرف الصقر يشويه. |

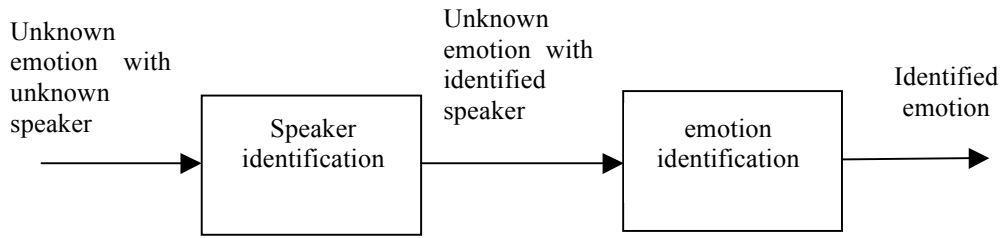

**Figure 1.** Block diagram of the overall proposed two-stage emotion recognizer

TABLE II.  EMOTION RECOGNITION ACCUACY BASED ON THE ONE-STAGE APPROACH.

| Emotion | Emotion Recognition Accuracy (%) |
|---|---|
| Neutral | 84.2 |
| Happy | 63.4 |
| Sad | 61.5 |
| Disgust | 55.9 |
| Angry | 40.2 |
| Fear | 63.2 |

TABLE III.  EMOTION RECOGNITION ACCUACY BASED ON THE TWO-STAGE APPROACH.

| Emotion | Emotion Recognition Accuracy (%) |
|---|---|
| Neutral | 90.4 |
| Happy | 70.1 |
| Sad | 66.7 |
| Disgust | 61.8 |
| Angry | 48.6 |
| Fear | 67.6 |